\documentclass[11pt]{article}

\usepackage[margin=1in]{geometry}
\usepackage{caption}
\usepackage{graphicx}
\usepackage{array}
\usepackage{booktabs}
\usepackage{tabularx}
\usepackage{xurl}
\usepackage{hyperref}
\usepackage{amsmath}
\usepackage{multirow}
\usepackage{makecell}
\usepackage{float}
\usepackage{IEEEtrantools}
\usepackage{xcolor}
\usepackage[section]{placeins}

\title{Scale-Dependent Input Representation and Confidence Estimation for LLMs in Materials Property Prediction}

\author{
Shuichiro Ozawa$^1$, Izumi Takahara$^1$, Teruyasu Mizoguchi$^1$ \\
$^1$Institute of Industrial Science, The University of Tokyo, Tokyo 153-8505, Japan \\
}

\begin{document}
\bstctlcite{IEEEexample:BSTcontrol}

\maketitle

\begin{abstract}

Large language models (LLMs) are increasingly applied to materials science. However, the relationship between prediction accuracy, input representation, and model scale remains unclear, and reliable methods for assessing prediction confidence have not yet been established. In this study, we fine-tune two Llama models of different scales (1B and 8B) using low-rank adaptation (LoRA) on an inorganic crystal structure dataset. We systematically evaluate five input representations, namely chemical composition, crystal summary, local environment description, full text description, and crystallographic information files (CIF), for formation energy and bandgap prediction. Our results show that the optimal input representation depends on model scale. The 1B model performs better with compact representations, whereas the 8B model maintains high accuracy even with longer natural-language descriptions and CIF inputs. Across both model scales, crystal summaries that include space-group information consistently outperform composition-only inputs, indicating that symmetry information serves as a robust and informative feature. We further analyze the relationship between prediction error and the mean negative log-likelihood (mean NLL) of tokens corresponding to predicted numerical values. While no clear correlation is observed in base models, fine-tuned models exhibit a consistent trend in which lower mean NLL corresponds to smaller prediction errors. This result suggests that mean NLL can serve as a practical confidence indicator without requiring additional training. These findings demonstrate that both input representation and model scale play critical roles in LLM-based materials property prediction, and that mean NLL provides an effective and computationally efficient measure of prediction confidence.

\end{abstract}

\section{Introduction}

Large language models (LLMs) have achieved substantial advances in natural language processing\cite{Brown2020-nt, Grattafiori2024-td} and are now applied to a broad area of scientific domains beyond language, including materials discovery\cite{Jia2024-mr, Takahara2025-ju}, literature mining\cite{Venugopal2024-fn, Shetty2023-kk}, and materials property prediction\cite{Niyongabo_Rubungo2025-cc, Tang2025-oy, Hu2025-wj, Niyongabo-Rubungo2025-vd}. Through pre-training on extensive chemical and materials science literature, LLMs encode considerable domain knowledge\cite{Gupta2022-zr, Taylor2022-ht, Mishra2024-mi} and may be capable of modeling structure-property relationships within a chemical context\cite{Korolev2023-sb, Jacobs2024-jc, Niyongabo_Rubungo2025-cc}. 

Despite the rapid growth in the application of LLMs, two key questions remain unresolved in materials science. Although recent benchmarks have begun to evaluate text- and crystallographic information file (CIF)-based representations for property prediction \cite{Alampara2025-so, Niyongabo-Rubungo2025-vd}, it remains unclear how the optimal input representation depends on model scale and fine-tuning conditions under controlled settings. Prior work has explored diverse approaches. Composition-based models such as Roost \cite{Goodall2020-ia} and CrabNet \cite{Wang2021-rj} demonstrate that accurate prediction is possible without explicit structural information, while other methods incorporate structural and contextual information through hybrid or language-based models, including CrysMMNet \cite{Das2023-de}, LLM-Prop \cite{Niyongabo_Rubungo2025-cc}, and ElatBot \cite{Liu2025-tw}. In addition, LLMs have been shown to process structured representations such as CIF and even generate crystal structures autoregressively \cite{Gruver2024-xf, Antunes2024-ri}. These studies highlight that materials can be represented in multiple ways, from chemical composition to natural-language descriptions and structured formats such as CIF. However, prior work adopts different input formats and model sizes, making it difficult to systematically determine how representation choice interacts with model scale and fine-tuning. As a result, this relationship remains insufficiently understood. 

\begin{figure}[!t]
    \centering
    \includegraphics[width=\linewidth]{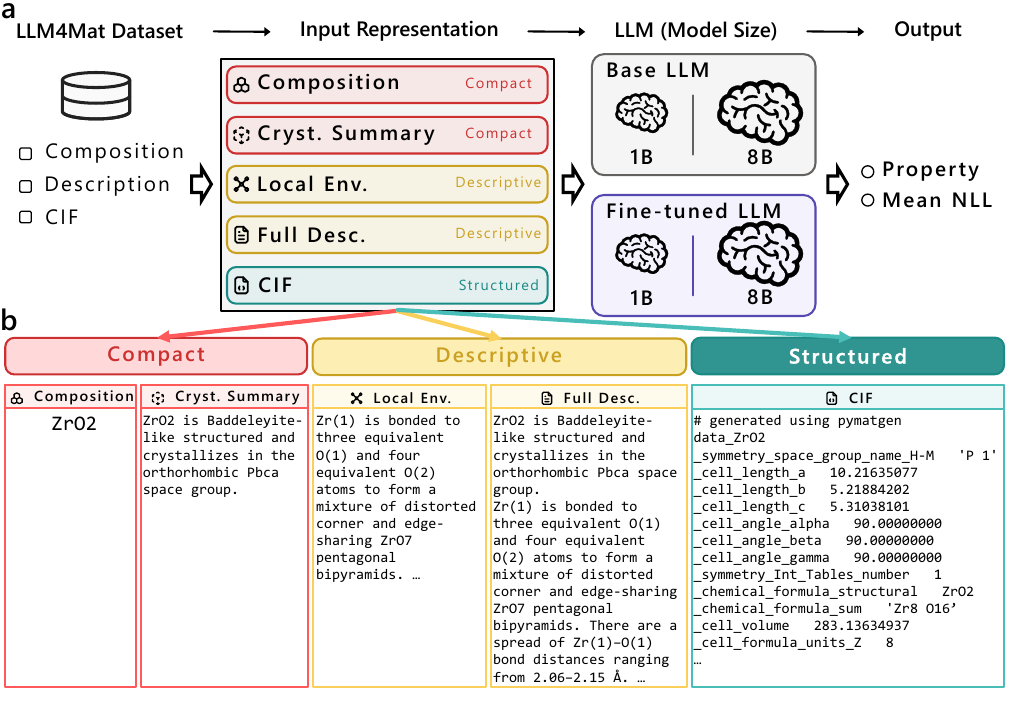}
    \caption{Experimental overview and examples of input representations. (a) Experimental pipeline in which five input representations are constructed from the LLM4Mat-Bench dataset and fed into 1B and 8B Llama models (base LLM and fine-tuned LLM) to output property values and mean NLL. Input representations are grouped into three categories by information content: Compact (red: Composition and Crystal Summary), Descriptive (yellow: Local Environment and Full Description), and Structured (green: CIF). (b) Concrete examples of each input representation using ZrO$_2$.}
    \label{fig:m1}
\end{figure}

Another unresolved question is how to assess the confidence of LLM-generated property predictions. Uncertainty quantification (UQ) is essential in ML-guided materials discovery, where overconfident predictions can misdirect costly experimental validation \cite{Varivoda2023-lt}. Existing UQ methods for graph-neural-network-based property prediction, such as ensemble methods, Monte Carlo dropout, deep evidential regression, and conformal prediction \cite{Dietterich2000-ba, Seoh2020-bb, Amini2020-ag, Angelopoulos2020-zl}, have been systematically benchmarked \cite{Varivoda2022-sd} but typically require additional modeling or computational overhead. In contrast, LLMs naturally provide token-level probabilities, enabling direct computation of negative log-likelihood (NLL). However, text-generation-based prediction can produce physically unrealistic values \cite{Ji2022-xu}, and without a reliable confidence metric, such errors cannot be filtered. Although NLL has been used as a confidence indicator in natural language processing \cite{Geng2024-qo, Vashurin2025-rc}, its applicability to materials property prediction remains unclear.

This study addresses two open questions in applying LLMs to materials science: input representation design and prediction confidence assessment. We investigate these questions in the context of materials property prediction, where prior work has shown that fine-tuning enables reasonable predictive accuracy \cite{Qu2024-tu, Choudhary2024-bi}. We conduct systematic experiments using five input representations and two Llama models of different scales (1B and 8B) to quantify the effects of input representation, model scale, and fine-tuning on predictive performance. We further examine the relationship between the mean NLL of predicted numerical tokens and prediction error, evaluating its utility as a confidence indicator in fine-tuned models.

The main contributions of this study are twofold. First, we demonstrate that the optimal input representation depends on both model scale and fine-tuning status through systematic experiments across input representations, model scales, and training conditions, providing practical guidance for input design in LLM-based materials property prediction. Second, we show that the mean NLL of numerical tokens in fine-tuned models correlates with prediction error, indicating its effectiveness as a confidence indicator.

\section{Methods}

\subsection{Dataset and input representations}

We use the inorganic crystal dataset provided by LLM4Mat-Bench \cite{Niyongabo-Rubungo2025-vd}, which is constructed from the Materials Project \cite{Horton2025-at}. The benchmark includes multiple materials properties for each sample and supports three types of input representations: chemical composition, natural-language descriptions of crystal structure based on Robocrystallographer \cite{Ganose2019-qg}, and CIF. We focus on formation energy per atom and bandgap as prediction targets and adopt the official train, validation, and test splits for model development and evaluation. The distributed dataset contains a subset of test samples with inconsistencies between material IDs and their associated property values or structural data. To ensure data integrity for CIF-based experiments, we retain the original sample IDs and property labels but re-fetch the corresponding crystal structures from the Materials Project to reconstruct the CIF inputs.

Five input representations are constructed for each sample. Composition encodes only the chemical formula. Crystal Summary consists of the leading sentence of the natural-language description, which includes composition and space-group information. Local Environment comprises the remaining description text excluding this sentence, while Full Description uses the complete text. CIF corresponds to the raw CIF string. The dataset contains 125,825 training, 10,000 validation, and 10,318 test samples. After excluding samples with missing input fields or target values, the valid sample counts are 125,822 (train), 10,000 (val), and 10,318 (test) for formation energy, and 125,098 (train), 9,945 (val), and 10,259 (test) for bandgap. For CIF-based experiments, crystal structures are additionally re-fetched from the Materials Project to correct inconsistencies in the distributed dataset, resulting in 10,223 and 10,166 valid test samples for formation energy and bandgap, respectively. Further details are provided in the Supplementary~\ref{app:data-prepro}.

A unified chat-format input is used across all experiments. The system prompt instructs the model to predict the target property as a materials scientist and to generate the output in JSON format. The user prompt provides the material representation corresponding to each input modality along with the target property name. Training labels are given in a single-key JSON format, with units of eV/atom for formation energy per atom and eV for bandgap. The complete prompt template is provided in Supplementary Table~\ref{tab:system_prompt}.

\subsection{Models and fine-tuning}

We use Llama-3.2-1B-Instruct (1B) and Llama-3.1-8B-Instruct (8B) \cite{Grattafiori2024-td} as base models. Both share the same auto-regressive transformer architecture but differ in parameter count and pre-training recipe. In the base setting, the models are used directly for inference without task-specific training. Fine-tuned models are trained using low-rank adaptation (LoRA) \cite{Hu2021-hj}.

LoRA is applied to the query and value projection layers of the attention modules, with rank $r = 32$, scaling factor $\alpha = 64$, and dropout $= 0.05$. The rank is selected based on an ablation study on the Llama 8B model for the formation energy task with Crystal Summary input, using $r = 4, 8, 16,$ and $32$ (see Supplementary Fig.~\ref{fig:rank}). Although the mean absolute error (MAE) decreases with increasing rank, the improvement plateaus near $r = 32$, which is therefore used in all experiments. Models are trained for 6 epochs with a learning rate of $1 \times 10^{-4}$, a per-device batch size of 1, gradient accumulation steps of 32, and a maximum input length of 8,192 tokens. The best checkpoint is selected based on validation loss.

Training is repeated across 60 conditions (3 seeds combined with task, model, and input representation), and convergence is observed by epoch 6, with validation loss improvements below 0.7\% between epochs 5 and 6. Training and inference times are provided in Supplementary Fig.~\ref{fig:time}. Training requires 25–42 hours for the 1B model and 38–120 hours for the 8B model, while inference on the test set takes 2–16 hours. Compared to CGCNN \cite{Xie2017-cq} (approximately 60 seconds for $\sim$10,000 samples), the higher inference cost reflects the overhead of auto-regressive generation.

The training objective is the cross-entropy loss over the response token sequence $y = (y_1, \ldots, y_T)$:
$$\mathcal{L} = -\frac{1}{T}\sum_{t=1}^{T} \log p_\theta(y_t \mid y_{<t},\ x)$$
, where $T$ is the total number of response tokens, $x$ is the input prompt, 
and $p_\theta$ denotes the model's token probability distribution. To assess reproducibility, training was repeated under three random seeds: 43, 44, and 45.

\subsection{Inference and uncertainty estimation}

All inference is performed using greedy decoding, with a maximum generation length of 512 tokens and a maximum input length of 8,192 tokens.

Calibration, defined as the degree to which predicted probabilities reflect true correctness rates, has been widely studied in deep learning as a measure of prediction confidence \cite{Pakdaman-Naeini2015-kj, Guo2017-sh}. We use the mean NLL computed from token probabilities at generation time as a confidence indicator and evaluate its effectiveness through exploratory analysis.
The NLL of token $y_t$ is the negative logarithm of the probability assigned by the model:

  $$\text{NLL}(y_t) = -\log p_\theta(y_t \mid y_{<t},\ x).$$
  
A lower NLL indicates higher model confidence in that token. The joint probability of a generated sequence is the product of per-token probabilities, which decreases exponentially with sequence length and is not directly comparable across sequences of different lengths. 

Taking the negative logarithm converts the product of token probabilities into a sum, and averaging over tokens normalizes for sequence length, yielding a length-invariant confidence measure. However, mean NLL computed over the full response still varies with sequence length, as longer responses tend to include more low-uncertainty tokens. To address this, we restrict the averaging to tokens corresponding to the predicted numerical value. Specifically, we consider the set $J$ of token indices whose character spans overlap with the integer part of the numerical string (including the sign and decimal point), and compute the mean NLL as

\begin{equation}
\text{Mean NLL} = -\frac{1}{|J|} \sum_{t \in J} \log p_\theta(y_t \mid y_{<t}, x),
\end{equation}

where $|J|$ denotes the number of tokens in the set.  This design is motivated by the greater reliability of the integer part as a confidence indicator. Fractional digit tokens are selected from a larger candidate set due to tokenization, which tends to increase their NLL regardless of prediction quality and introduces noise unrelated to model confidence. In contrast, the integer part involves fewer tokens with less ambiguous selection, making its mean NLL a more reliable proxy for uncertainty. We also define an alternative measure, Full Numeric NLL, in which $J$ spans all tokens of the numerical string, including fractional digits. A comparison of the two measures is provided in Supplementary Figs.~\ref{fig:s_nll_base}, \ref{fig:s_nll_ft}.

\subsection{Evaluation}

A generation is considered a parse success if a numerical value can be extracted from the first JSON object in the output, and the fraction of such cases across the test set is reported as the parse success rate. As shown in Supplementary Table~\ref{tab:parse_success}, the parse success rate exceeds 85\% even for base models (except for the 1B Local Environment condition) and improves to over 89\% across all conditions after fine-tuning, indicating reliable adherence to the specified JSON format.

For successfully parsed samples, prediction accuracy is evaluated using mean absolute error (MAE) between predicted and true values. Root mean square error (RMSE) is additionally reported in Supplementary Table~\ref{tab:rmse}. For comparison, we use CGCNN \cite{Xie2017-cq} as a GNN baseline, trained from scratch on the LLM4Mat-Bench dataset with standard hyperparameters, including a hidden dimension of 128, batch size of 256, three message-passing layers, a learning rate of 0.01, a cutoff radius of 8.0~\AA, 12 nearest neighbors, and 500 training epochs.

\section{Results and Discussion}

\subsection{Effects of input representation and model scale on property prediction}

\begin{figure}[!t]
    \centering
    \includegraphics[width=\linewidth]{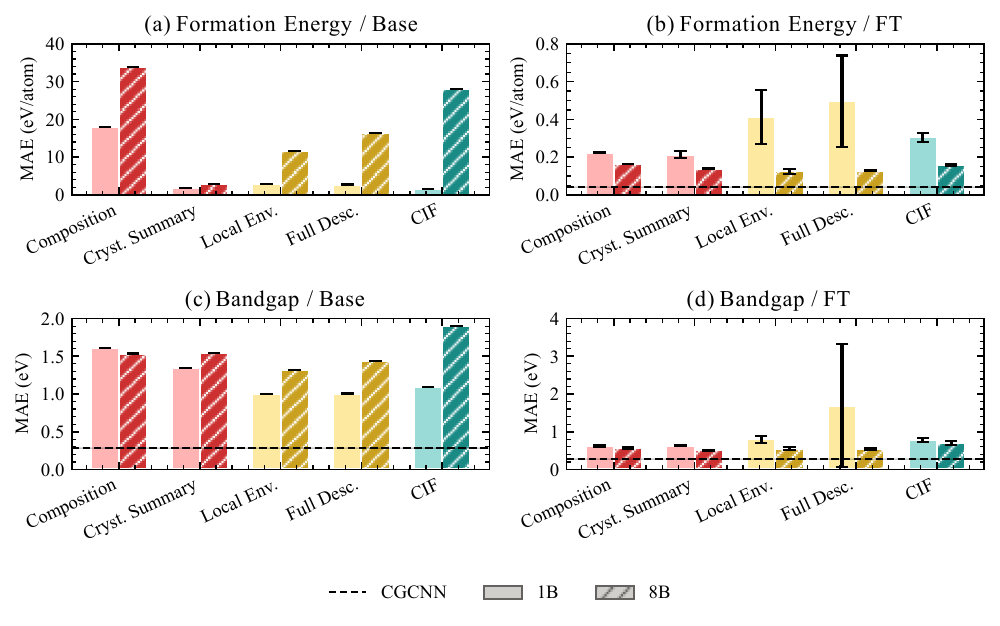}
    \caption{MAE for five input representations under base and fine-tuned model conditions. Four panels are shown: (a) formation energy, base model; (b) formation energy, fine-tuned; (c) bandgap, base model; (d) bandgap, fine-tuned. Input representations are color-coded: Composition and Crystal Summary (compact inputs) in red, Local Environment and Full Description (natural-language descriptions) in yellow, and CIF in green. Light and dark fills (with hatching) denote 1B and 8B models, respectively. Note that the y-axis in (a) extends to approximately 40 eV/atom to accommodate large base model errors, whereas (b) is scaled to 0.8 eV/atom.}
    \label{fig:m2}
\end{figure}

We first evaluate property prediction accuracy across all input representations and model conditions. Figure~\ref{fig:m2} shows the MAE for all conditions, with numerical values summarized in Table~\ref{table:m1}. Formation energy and bandgap results are presented in Fig.~\ref{fig:m2}(a,b) and Fig.~\ref{fig:m2}(c,d), respectively.

Under the base model conditions (Table~\ref{table:m1}), representations such as Crystal Summary yield lower MAE than Composition for formation energy (Fig.~\ref{fig:m2}(a)). However, since the base models are not optimized for numerical prediction, this apparent improvement does not reflect meaningful physical understanding. In particular, the 1B base model consistently outputs values near 0 eV for natural-language inputs other than Composition, indicating a failure to produce meaningful predictions. The resulting lower MAE arises from this trivial output pattern rather than improved modeling of structural information, as ground truth values often have small magnitudes. These results highlight that task-specific fine-tuning is necessary to properly assess the impact of input representations on prediction accuracy.

\begin{table}[htbp]
    \centering
    \small
    \caption{MAE for base and fine-tuned models on the formation energy task. Values are reported as mean $\pm$ standard deviation across three random seeds. CGCNN (GNN baseline) and encoder-based LLMs (LLM-Prop, MatBERT) are included for comparison. Units: eV/atom. Encoder-based model with a regression head; direct comparison with the decoder-based models is not straightforward, and values are provided for reference only.}
    \label{table:m1}
    \begin{tabular}{llccccc}
    \toprule
     & Modality & 1B & 8B & CGCNN & LLM-Prop & MatBERT \\
    \midrule
    \multirow{5}{*}{Base}
     & Composition      & 18.05 & 33.99 & - & - & - \\
     & Cryst. Summary   & 1.743 & 2.912 & - & - & - \\
     & Local Env.       & 2.889 & 11.57 & - & - & - \\
     & Full Desc.       & 2.728 & 16.45 & - & - & - \\
     & CIF              & 1.515 & 28.10 & - & - & - \\
    \midrule
    \multirow{5}{*}{Fine-tuned}
     & Composition      & $0.223 \pm 0.004$ & $0.164 \pm 0.001$ & - & 0.134 & 0.123 \\
     & Cryst. Summary   & $0.213 \pm 0.018$ & $0.137 \pm 0.002$ & - & - & - \\
     & Local Env.       & $0.411 \pm 0.143$ & $0.124 \pm 0.012$ & - & - & - \\
     & Full Desc.       & $0.497 \pm 0.242$ & $0.128 \pm 0.003$ & - & 0.063 & 0.084 \\
     & CIF              & $0.305 \pm 0.025$ & $0.158 \pm 0.005$ & $0.042 \pm 0.001$ & 0.070 & 0.091 \\
    \bottomrule
    \end{tabular}
\end{table}

Upon fine-tuning, predictive accuracy improved substantially across all conditions, reaching levels suitable for practical property prediction (Fig.~\ref{fig:m2}(b),(d)). Clear trends emerged regarding the relationship between model scale and the optimal input representation, as detailed in Table~\ref{table:m1}. For the 1B fine-tuned model, shorter and more abstract representations such as Composition and Crystal Summary achieved higher accuracy. As inputs became more detailed (e.g., Local Environment and Full Description), the MAE increased, accompanied by larger standard deviations across different random seeds. This suggests that smaller models lack sufficient capacity to map complex, long-form text descriptions to precise physical properties, where excessive information may lead to training instability.

In contrast, for the formation energy task, the 8B fine-tuned model exhibited the opposite trend: accuracy improved as input descriptions became more detailed. For the formation energy task, the 8B fine-tuned model consistently recorded its lowest MAE when provided with the Full Description. This indicates that larger models can leverage their superior natural language understanding capabilities, gained during pre-training, to effectively extract nuanced structural features from descriptive text for property prediction. These results underscore the importance of selecting an appropriate level of information abstraction based on the model's scale.

This scale-dependent capability was also evident in the processing of the CIF format. While the 1B fine-tuned model struggled to derive useful structural patterns from CIF strings, the 8B fine-tuned model successfully interpreted this structured data, performing comparably to or marginally better than  the Composition-only baseline. Nevertheless, even for the 8B fine-tuned model, CIF-based accuracy did not reach the performance levels of the natural-language-based Full Description. This implies that natural-language descriptions of crystal structures are more compatible with the LLM's internal representations, which are primarily optimized for natural language, than raw coordinate data.

A detailed analysis of fine-tuned model results highlights the critical role of symmetry information. Crystal Summary, which incorporates space-group information, outperforms Composition across all model scales and tasks. While chemical composition alone cannot distinguish between polymorphs, space-group information provides a clear disambiguation signal. More broadly, symmetry descriptors such as space-group labels may help activate crystallographic knowledge embedded in the LLM during pretraining, offering a compact and information-rich representation for property prediction. To further examine this effect, we evaluate the model on SiO$_2$ polymorphs (cristobalite, quartz, and tridymite). As shown in Table~\ref{table:m2}, representations that include symmetry information enable the model to distinguish structural differences and improve prediction accuracy.

\begin{table}[htbp]
    \centering
    \small
    \caption{Predicted formation energies for SiO$_2$ polymorphs and corresponding MAE for 1B and 8B fine-tuned models. Units: eV/atom.}
    \label{table:m2}
    \begin{tabular}{llcccc}
    \toprule
    Model Scale & Modality 
    & \makecell{Cristobalite \\ ($I4/mmm$)}
    & \makecell{Quartz \\ ($P6_222$)}
    & \makecell{Tridymite \\ ($P6/mmm$)}
    & MAE \\
    \midrule
    \multirow{5}{*}{1B}
     & Composition      & -3.1511 & -3.1511 & -3.1511 & 0.1041 \\
     & Cryst. Summary   & -3.2552 & -3.1716 & -3.2516 & \textbf{0.0291} \\
     & Local Env.       & -2.9729 & -1.9340 & -2.0453 & 0.9378 \\
     & Full Desc.       & -1.3339 & -1.0000 & -2.6582 & 1.5912 \\
     & CIF              & -3.2970 & -3.3098 & -3.0545 & 0.0992 \\
    \midrule
    \multirow{5}{*}{8B}
     & Composition      & -3.2539 & -3.2539 & -3.2539 & 0.0013 \\
     & Cryst. Summary   & -3.2551 & -3.2518 & -3.2552 & \textbf{0.0012} \\
     & Local Env.       & -3.2143 & -3.2416 & -3.1694 & 0.0468 \\
     & Full Desc.       & -3.2539 & -3.2573 & -3.2526 & 0.0025 \\
     & CIF              & -3.1728 & -3.2133 & -2.7445 & 0.2117 \\
    \midrule
    \multicolumn{2}{l}{Ground Truth} & -3.2557 & -3.2545 & -3.2555 & 0.0000 \\
    \bottomrule
    \end{tabular}
\end{table}

On the other hand, the benefits of utilizing more detailed input representations (i.e., adding extended natural-language descriptions) depend on the physical nature of the target property. As discussed earlier, for formation energy prediction, detailed representations such as Full Description substantially improved the 8B fine-tuned model's accuracy. In contrast, for the bandgap task, the benefits of adding detailed descriptions were limited. This contrast likely reflects the nature of each property. Namely, formation energy benefits directly from detailed structural and compositional information encoded in natural-language descriptions, whereas bandgap is more strongly governed by electronic structure near the Fermi level. Although elemental composition provides indirect cues, such as electronegativity differences, that partially capture bonding character, the additional structural detail in longer descriptions does not further improve prediction, suggesting that the relevant electronic information lies beyond what current crystal descriptions encode. 

Finally, comparing our 8B fine-tuned model with existing benchmarks, the MAE did not surpass specialized models like CGCNN (0.042~eV/atom) or encoder-based LLMs like LLM-Prop\cite{Niyongabo_Rubungo2025-cc} (0.063~eV/atom)\cite{Niyongabo-Rubungo2025-vd} and MatBERT\cite{Shetty2023-kk} (0.123 eV/atom)\cite{Niyongabo-Rubungo2025-vd}. This gap stems from fundamental differences in architecture: GNNs explicitly encode geometric relationships, and encoder-based models directly optimize for regression through a dedicated prediction head. In contrast, auto-regressive decoder models like Llama treat numerical prediction as a discrete token generation task, which poses inherent challenges for continuous value regression. Nonetheless, the flexibility of natural-language input design and the potential for multi-task application without task-specific architecture represent practical advantages that complement specialized models in materials science workflows. 

\subsection{Confidence estimation using negative log-likelihood}

The results above demonstrate that input representation and model scale jointly shape prediction accuracy. A remaining practical question is whether the model's own output distribution can signal when its predictions are reliable. In this section, we examine whether mean NLL, a token-probability-based confidence measure established in NLP, can serve as a confidence indicator for LLM-based materials property prediction.

Under the base model conditions, no meaningful correlation was observed between mean NLL and prediction error; large errors frequently appeared even at low mean NLL values, indicating that the output probabilities do not reflect prediction reliability. This is consistent with the expectation that, without task-specific fine-tuning, the model's generation probabilities reflect pretraining biases rather than the material–property relationship (see Supplementary Fig.~\ref{fig:s_scatter_base}).

\begin{figure}[ht]
    \centering
    \includegraphics[]{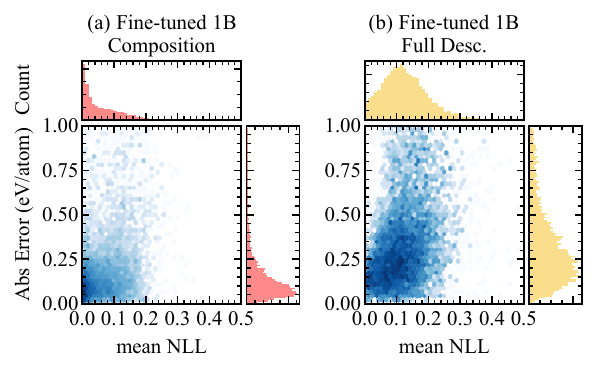}
    \caption{Scatter plots of mean NLL versus absolute error for the fine-tuned 1B model on the formation energy task, using (a) Composition input and (b) Full Description input. Each panel includes a scatter plot with marginal histograms of mean NLL (top) and absolute error (right).}
    \label{fig:m3}
\end{figure}

For the 1B fine-tuned model, we compare the Composition and Full Description conditions (Fig.~\ref{fig:m3}). Under Composition input, mean NLL values are concentrated in the range of 0.0–0.2, and absolute errors in this region remain small. Under Full Description input, mean NLL values are more broadly distributed up to approximately 0.3, with larger absolute errors also present. This broader distribution is consistent with the increased training instability observed for the 1B model under long-context inputs, suggesting that input complexity affects not only prediction accuracy but also confidence calibration. Nevertheless, a tendency for lower mean NLL to correspond to smaller absolute error is visually apparent in both conditions. A similar trend was observed for the 8B model (see Supplementary Fig.~\ref{fig:s_scatter_ft}). These results suggest that, in fine-tuned models, mean NLL can function as a confidence indicator for materials property prediction.

We therefore introduce NLL filtering: retaining only samples whose mean NLL falls below a given threshold, and discarding the rest as low-confidence predictions. Selective prediction of this kind has been shown to be effective for classification and question-answering tasks in NLP \cite{Xin2021-ce}, and the present work extends this framework to regression-based property prediction. By varying the threshold continuously and tracking the resulting MAE, we quantify how much accuracy can be gained by abstaining from uncertain predictions—at the cost of reduced coverage, defined as the fraction of samples for which a prediction is made.  This analysis is exploratory; in practice, the threshold should be determined on a held-out validation set before deployment. 

\begin{figure}[t]
    \centering
    \includegraphics[width=\linewidth]{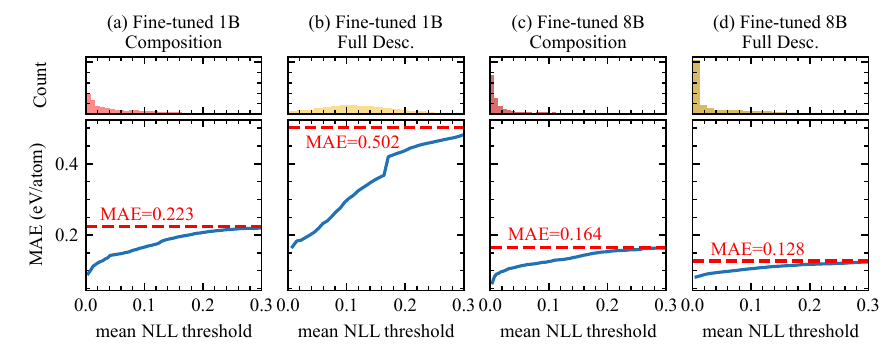}
    \caption{Change in MAE with NLL filtering for fine-tuned models. Panels (a) and (b) compare Composition and Full Description inputs for the 1B model; panels (c) and (d) show the corresponding comparison for the 8B model. Each panel shows a count histogram (top) and a line plot of MAE versus mean NLL threshold (bottom); the red dashed line indicates the unfiltered MAE (baseline).}
    \label{fig:m4}
\end{figure}

Figure \ref{fig:m4} shows how MAE changes with NLL filtering for the formation energy task. Each panel shows a count histogram (top) and a line plot of MAE versus mean NLL threshold (bottom), with a red dashed line indicating the unfiltered MAE. From left to right: 1B Composition, 1B Full Description, 8B Composition, and 8B Full Description. Results for all conditions are shown in Supplementary Fig.~\ref{fig:s_nll_base} and \ref{fig:s_nll_ft}.

Overall, in the fine-tuned models, lowering the mean NLL threshold reduced MAE below the baseline across all conditions (Fig. \ref{fig:m4}), suggesting that NLL-based selective prediction tends to improve the reliability of materials property predictions. This trend was consistent across both model scales and input representations, indicating that mean NLL functions as a confidence indicator in fine-tuned models regardless of input complexity.

Fine-tuning is a prerequisite for this behavior: without it, the model lacks sufficient knowledge of the material–property relationship to produce calibrated output probabilities. These findings suggest that NLL filtering is a practical tool for improving prediction reliability in deployment, complementing the choice of input representation discussed in the preceding sections. However, its effectiveness depends on the alignment between model capacity and input complexity: as observed for the 1B model under Full Description input, higher input complexity can broaden the NLL distribution and reduce filtering precision. In practice, the filtering threshold should be determined on a held-out validation set, and NLL filtering should be applied in conjunction with careful input design.

To assess the robustness of this finding, we also computed an alternative measure, Full Numeric NLL, which averages over all tokens of the predicted numerical string including fractional digits, and present the results in Supplementary Fig.~\ref{fig:s_nll_base} and \ref{fig:s_nll_ft}. Unlike the integer-part mean NLL, Full Numeric NLL exhibited an increase in MAE at very low thresholds, meaning that samples assigned the highest confidence were not necessarily the most accurate. This discrepancy suggests that the model assigns unstable probabilities to fractional digit tokens: whereas the integer part is generated with relatively well-calibrated confidence, the model's token-level uncertainty becomes unreliable beyond the decimal point. These results support the use of integer-part mean NLL over Full Numeric NLL as a confidence indicator, and highlight that the choice of token scope is consequential for NLL-based selective prediction.

\section{Conclusion}

In this study, We investigated how input representation and model scale affect property prediction accuracy by fine-tuning Llama models (1B and 8B) using LoRA for formation energy per atom and bandgap prediction.

Our results show two key findings. First, the optimal input representation depends on model scale: the 1B model performs best with compact representations such as chemical composition and crystal summary, whereas the 8B model remains robust to more detailed inputs, including natural-language descriptions and CIF. Representations incorporating symmetry information, such as crystal summaries, consistently improve performance across model scales, indicating that symmetry information contributes to prediction accuracy.

Second, in fine-tuned models, the mean NLL of numerical tokens correlates with prediction error, while no such relationship is observed in base models. This finding extends token-based confidence estimation from NLP to materials property prediction and shows that fine-tuning is necessary for mean NLL to serve as a reliable confidence indicator. This analysis is exploratory and based on test-set observations; validation-based threshold selection is required for practical deployment.

This study has several limitations. The analysis is restricted to 1B and 8B models, and further investigation is required to generalize the observed scale dependence to a broader range of model sizes. In addition, only a limited set of input representations and target properties (formation energy and bandgap) are considered. The system prompt design follows LLM4Mat-Bench \cite{Niyongabo-Rubungo2025-vd} for consistency with prior work, and sensitivity to prompt variations is not explored. From a practical perspective, computational cost remains a key limitation. LLM inference requires 2–16 hours per test set of approximately 10,000 samples, compared to about 60 seconds for CGCNN, representing a difference of two to three orders of magnitude. The 8B model also requires 16–20 GB of GPU memory, whereas CGCNN can run with significantly lower memory or on CPU.

Future work includes extending the analysis to larger models (e.g., 70B scale), evaluating NLL-based filtering across a broader range of material properties, and establishing validation-based threshold selection for practical use. The findings reported here are specific to the Llama 3 series, and further validation is required to generalize them to other architectures. The insights from input representation design and NLL-based confidence assessment provide a practical starting point for applying LLMs to materials property prediction. The primary advantage of this approach lies in the flexibility of natural-language input design and its potential for multi-task application without task-specific architectures. As LLMs continue to scale and their pretraining distributions expand, the input-representation and confidence-estimation strategies identified here may become increasingly effective for materials informatics.

\section*{Code Availability}

The source code used in this study is available at \url{https://github.com/shu-ozawa/crystal-llm-representations}.
The fine-tuned models and training logs are available at \url{https://huggingface.co/ozashu/crystal-llm-representations}. 

\section*{Acknowledgment}

This study was supported by the NVIDIA Academic Grant Program. Computations were performed using NVIDIA A100 (80 GiB) GPUs provided through this program. This study was also supported by Japan Science and Technology Agency (JST) (Nos. JPMJAX24DB and JPMJBS2418), the Ministry of Education, Culture, Sports, Science and Technology (MEXT) (Nos. 24H00042 and 26K01205).

\bibliographystyle{IEEEtran}
\bibliography{reference}

\newpage
{\centering\section*{Supplementary Material}}
\appendix
\renewcommand{\thesection}{S\arabic{section}}
\setcounter{section}{0}
\setcounter{figure}{0}
\setcounter{table}{0}
\setcounter{equation}{0}
\renewcommand{\thefigure}{S\arabic{figure}}
\renewcommand{\thetable}{S\arabic{table}}
\renewcommand{\theequation}{S\arabic{equation}}

\section{Dataset Preprocessing for CIF-based Evaluation}
\label{app:data-prepro}

For CIF-based evaluation, we corrected inconsistencies in the distributed LLM4Mat-Bench test set. Although the test set contains 10,318 samples, the \texttt{cif\_structure} column is misaligned with the corresponding \texttt{material\_id} and metadata for approximately 4,858 samples. 

To address this issue, we retrieved CIF structures directly from the Materials Project using \texttt{material\_id} as the reference key. Among the 10,318 test materials, 95 structures could not be retrieved because their identifiers are no longer available in the current Materials Project database.

As a result, after excluding unretrievable structures and task-specific invalid entries, the CIF-based test set contains 10,223 valid samples for formation energy prediction and 10,166 valid samples for bandgap prediction.

\section{Prompt Template}

\begin{table}[H]
    \centering
    \caption{System prompt template used for material property prediction across all input modalities.}
    \label{tab:system_prompt}
    \begin{tabular}{p{0.95\linewidth}}
         \hline
         You are a materials scientist. \\
         Look at the \{input\_type\} of the given crystalline material and predict its property. \\ 
         The output must be in a json format. \\ For example: \{property\_name: predicted\_property\_value\}. \\
         Answer as precise as possible and in few words as possible. \\
         \hline
    \end{tabular}
\end{table}

\section{Hyperparameter Selection}

\begin{figure}[H]
    \centering
    \includegraphics{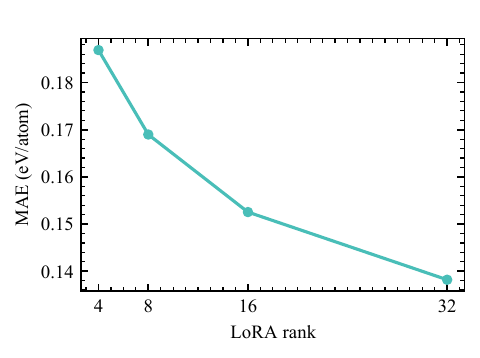}
    \caption{Effect of LoRA rank on validation MAE for the Llama-3.1-8B-Instruct model trained on the formation energy task with Crystal Summary input. Ranks r = 4, 8, 16, and 32 were evaluated. MAE decreases monotonically with increasing rank; the rate of improvement plateaus near r = 32, which was adopted for all subsequent experiments.}
    \label{fig:rank}
\end{figure}

\section{Computational Cost}

\begin{figure}[H]
    \centering
    \includegraphics[width=\linewidth]{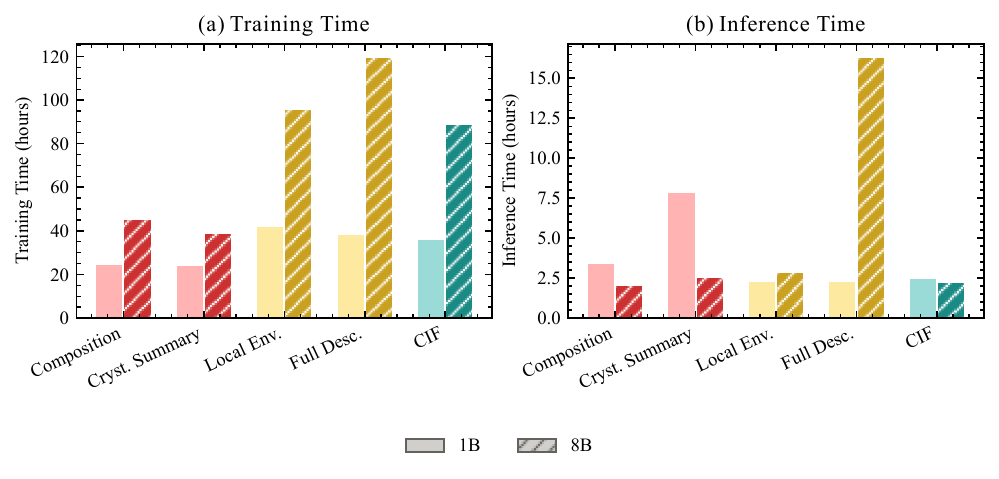}
    \caption{Training time per run (hours) and inference time on the test set for each combination of task, model size (1B and 8B), and input representation. Training time ranged from 25 to 42 hours for the 1B model and from 38 to 120 hours for the 8B model (longest: Full Description, 8B, approximately 120 hours per run). Inference on the test set required 2 to 16 hours per condition.}
    \label{fig:time}
\end{figure}

\section{Additional Quantitative Results}

\subsection{Parse Success Rate}

\begin{table}[H]
\centering
\small
\caption{Parse success rate (fraction of test samples from which a numerical value was successfully extracted) for all conditions across task (formation energy, bandgap), model size (1B, 8B), training condition (base, fine-tuned), and input representation (Composition, Crystal Summary, Local Environment, Full Description, CIF).}
\label{tab:parse_success}
\begin{tabular}{lllccccc}
\toprule
Task & Model & Setting & Composition & \makecell{Cryst. \\Summary} & Local Env. & Full Desc. & CIF \\
\midrule
\multirow{4}{*}{\makecell{Formation Energy \\ (eV/atom)}}
& \multirow{2}{*}{1B}
& Base       & 1.00 & 1.00 & 0.54 & 0.85 & 0.89 \\
&            & Fine-tuned & 1.00 & 1.00 & 0.97 & 0.94 & 0.97 \\
& \multirow{2}{*}{8B}
& Base       & 0.97 & 0.96 & 0.96 & 0.95 & 0.92 \\
&            & Fine-tuned & 1.00 & 1.00 & 0.98 & 0.98 & 0.91 \\
\midrule
\multirow{4}{*}{\makecell{Bandgap \\ (eV)}}
& \multirow{2}{*}{1B}
& Base       & 1.00 & 1.00 & 0.85 & 0.94 & 0.94 \\
&            & Fine-tuned & 1.00 & 0.99 & 0.98 & 0.92 & 0.92 \\
& \multirow{2}{*}{8B}
& Base       & 0.99 & 1.00 & 0.97 & 0.96 & 0.99 \\
&            & Fine-tuned & 1.00 & 1.00 & 0.98 & 0.98 & 1.00 \\
\bottomrule
\end{tabular}
\end{table}

\subsection{RMSE}

\begin{table}[H]
\centering
\small
\caption{RMSE for base and fine-tuned models across input modalities for formation energy and band gap prediction. Values are reported as mean $\pm$ standard deviation across three random seeds. Units: eV/atom for formation energy and eV for band gap. CGCNN is included as a GNN baseline for reference.}
\label{tab:rmse}
\begin{tabular}{llcccc}
\toprule
 & Modality & \multicolumn{2}{c}{Formation Energy} & \multicolumn{2}{c}{Bandgap} \\
 &  & 1B & 8B & 1B & 8B \\
\midrule
\multirow{5}{*}{Base}
 & Composition      & 46.04 & 105.66 & 19.78 & 1.89 \\
 & Cryst. Summary   & 4.66 & 24.62 & 17.16 & 1.96 \\
 & Local Env.       & 4.47 & 54.28 & 1.55 & 1.69 \\
 & Full Desc.       & 3.78 & 72.71 & 1.56 & 1.78 \\
 & CIF              & 1.84 & 106.65 & 1.69 & 13.92 \\
\midrule
\multirow{5}{*}{Fine-tuned}
 & Composition      & $0.491 \pm 0.012$ & $0.429 \pm 0.006$ & $1.190 \pm 0.039$ & $1.138 \pm 0.042$ \\
 & Cryst. Summary   & $0.405 \pm 0.012$ & $0.321 \pm 0.004$ & $1.195 \pm 0.015$ & $1.040 \pm 0.029$ \\
 & Local Env.       & $0.747 \pm 0.252$ & $0.233 \pm 0.015$ & $1.428 \pm 0.177$ & $1.104 \pm 0.082$ \\
 & Full Desc.       & $3.802 \pm 5.598$ & $0.232 \pm 0.002$ & $18.165 \pm 29.072$ & $1.089 \pm 0.055$ \\
 & CIF              & $0.522 \pm 0.028$ & $0.334 \pm 0.002$ & $1.350 \pm 0.051$ & $1.323 \pm 0.087$ \\
\midrule
CGCNN & -- & $0.097 \pm 0.005$ & $0.097 \pm 0.005$ & $0.564 \pm 0.003$ & $0.564 \pm 0.003$ \\
\bottomrule
\end{tabular}
\end{table}

\subsection{NLL--Error Scatter Plots}

\begin{figure}[H]
    \centering
    \includegraphics[width=0.9\linewidth]{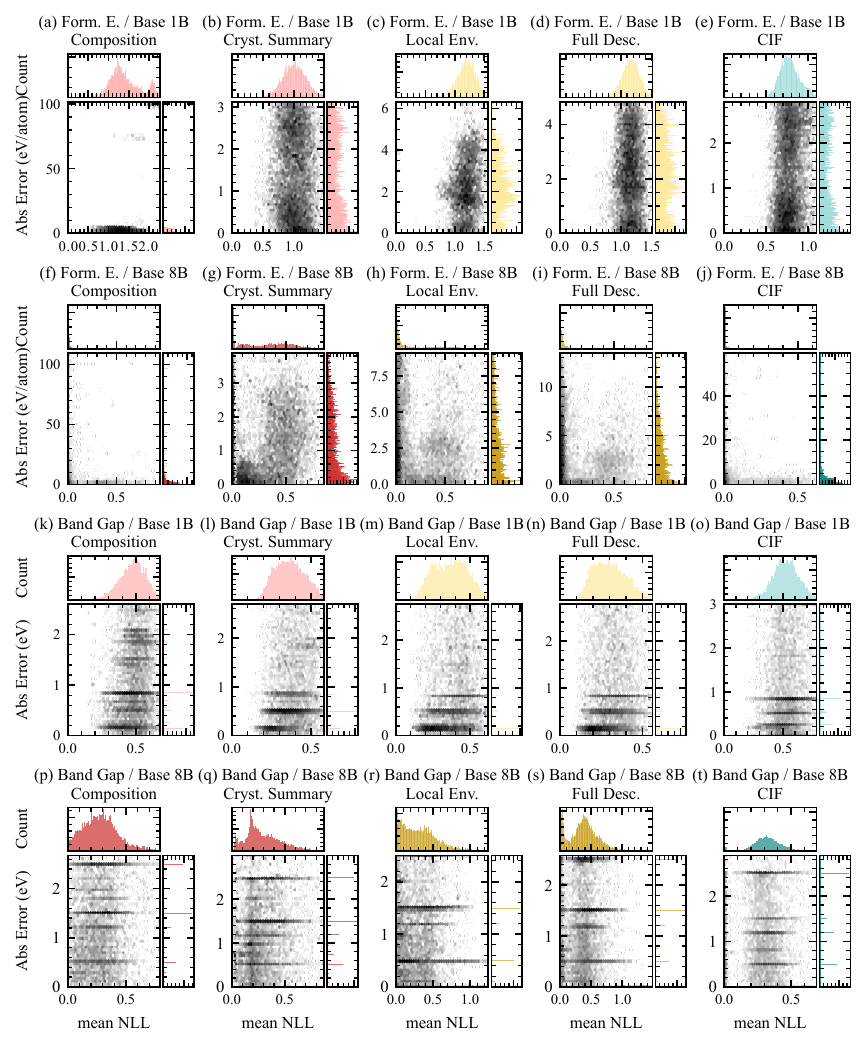}
    \caption{Scatter plots of mean NLL versus absolute error for all conditions (formation energy task). Each panel corresponds to one combination of model size (1B or 8B), training condition (base or fine-tuned), and input representation. Layout as in Fig.~\ref{fig:m3}: scatter plot with marginal histograms of mean NLL (top, pink) and absolute error (right, pink). Units: eV/atom.}
    \label{fig:s_scatter_base}
\end{figure}

\begin{figure}[H]
    \centering
    \includegraphics[width=0.9\linewidth]{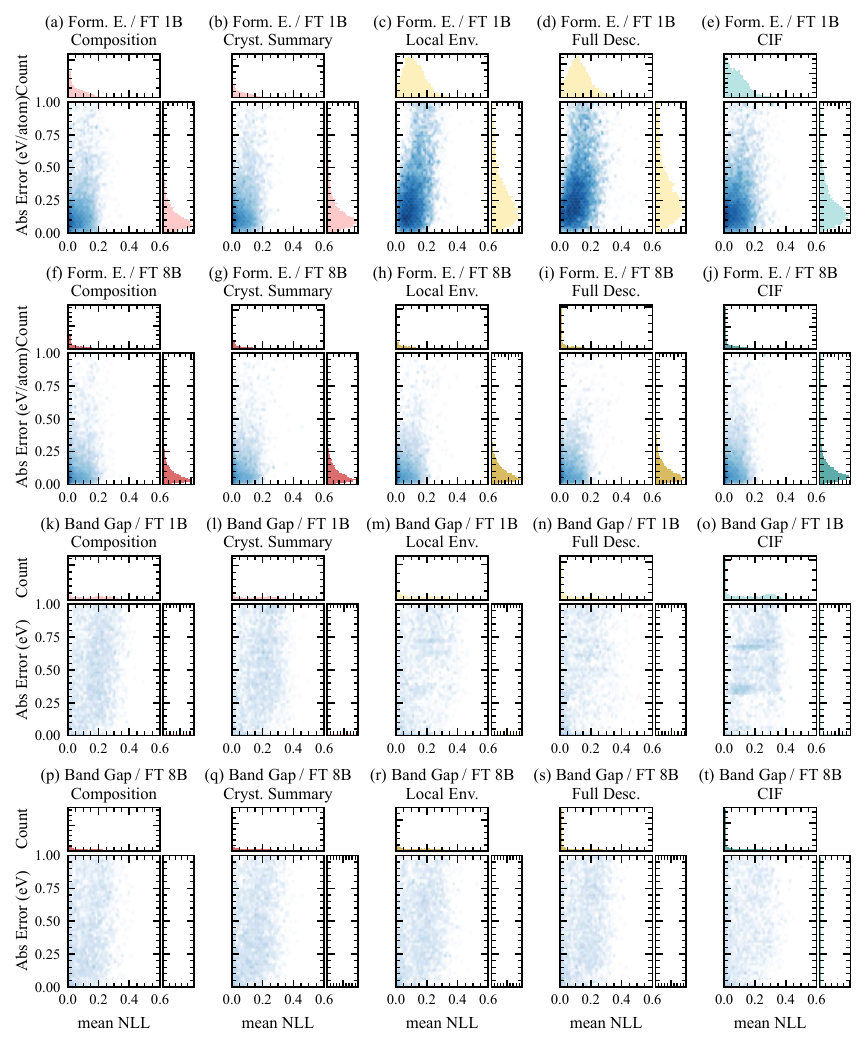}
    \caption{Scatter plots of mean NLL versus absolute error for all conditions (bandgap task). Layout as in Fig.~\ref{fig:s_scatter_base}. Units: eV.}
    \label{fig:s_scatter_ft}
\end{figure}

% Comparison of Integer Part and Full Numeric NLL
\begin{figure}[H]
    \centering
    \includegraphics[width=0.9\linewidth]{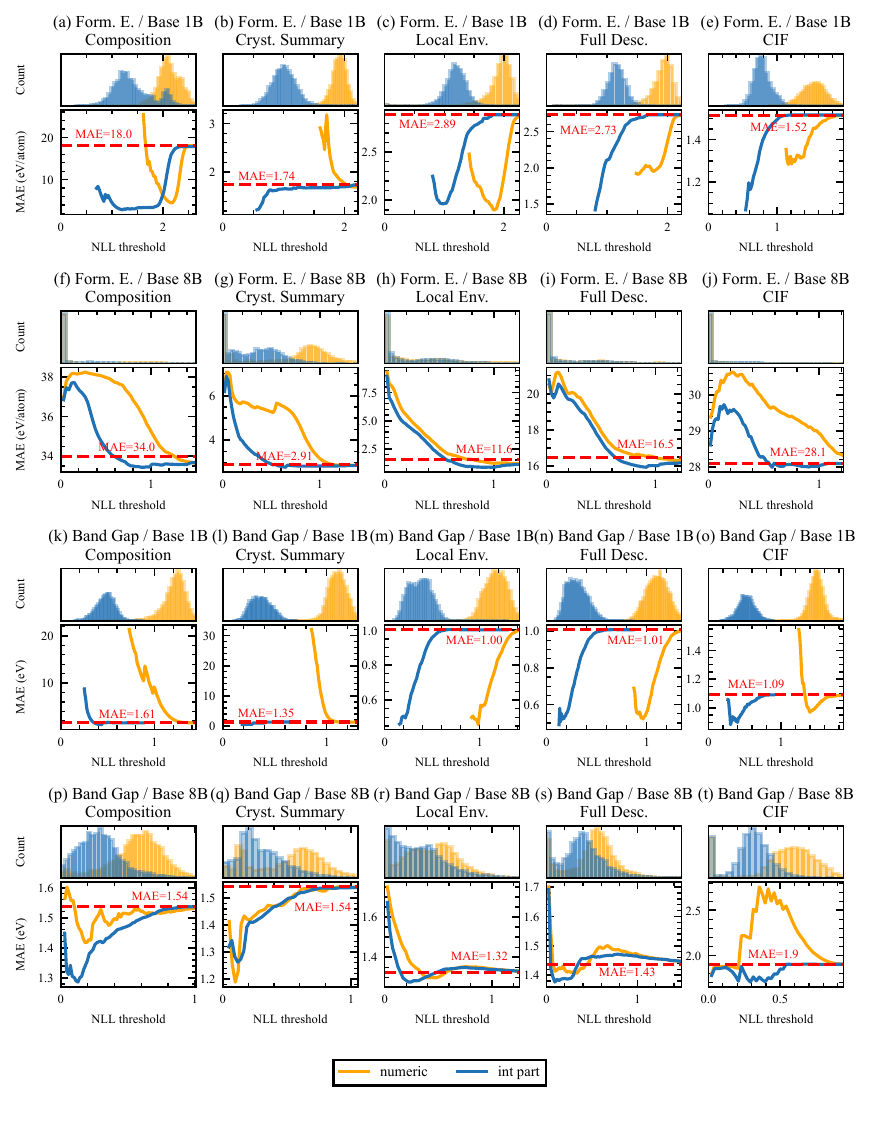}
    \caption{Effect of NLL-based confidence filtering on MAE for the base model. Rows: tasks (Formation Energy, Band Gap). Columns: input modalities (Composition, Cryst. Summary, Local Env., Full Desc., CIF). Top subplot: mean NLL distribution. Bottom subplot: MAE vs.\ NLL threshold $t$. Blue: Integer Part NLL; Orange: Full Numeric NLL. Red dashed line: unfiltered baseline MAE.}
    \label{fig:s_nll_base}
\end{figure}

\begin{figure}[H]
    \centering
    \includegraphics[width=0.9\linewidth]{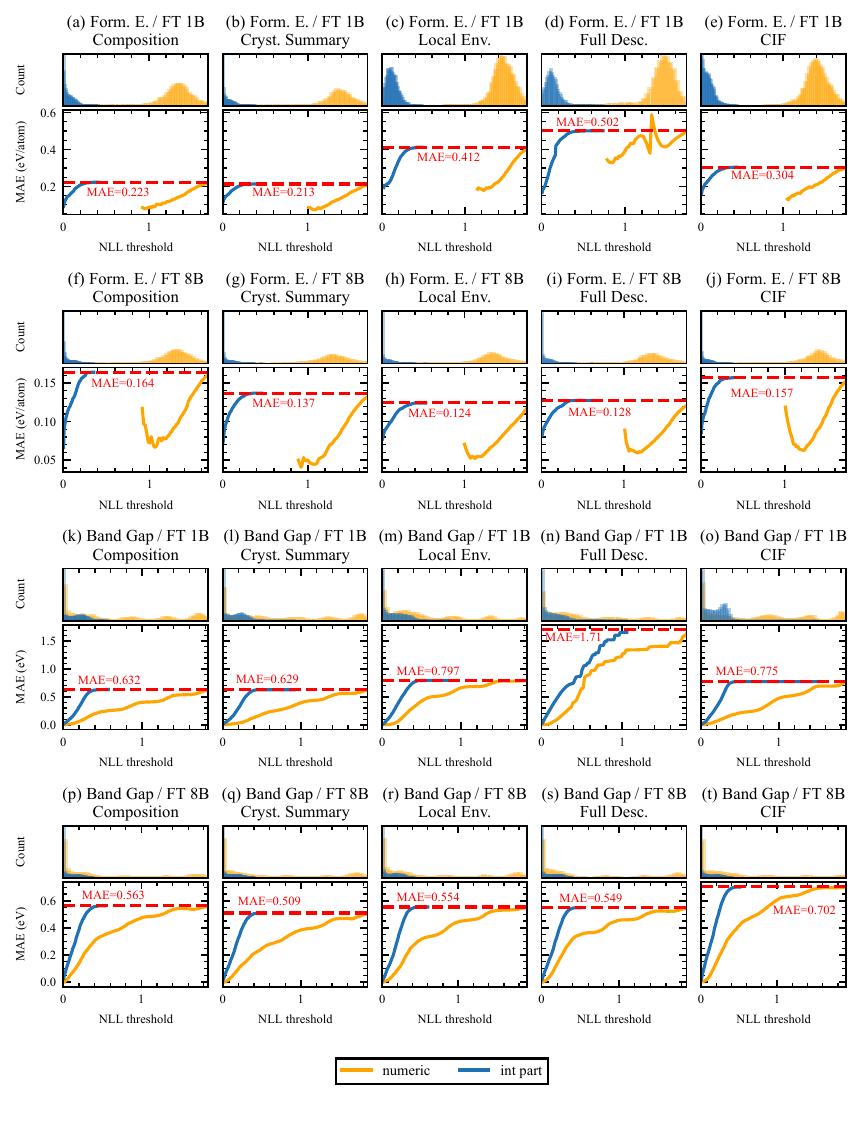}
    \caption{
    Same as Fig.~\ref{fig:s_nll_base}, but for the fine-tuned model.
    }
    \label{fig:s_nll_ft}
\end{figure}

\end{document}